\documentclass[12pt]{iopart}

\usepackage{graphicx}
\begin{document}

\title[]{Fabrication and measurements of hybrid Nb/Al
Josephson junctions and flux qubits with $\pi$-shifters}

\author{A~V~Shcherbakova$^{1}$, K G Fedorov$^{1,2}$,  K V Shulga$^{2,3}$, V~V~Ryazanov$^{2,3,4}$, V V Bolginov$^{2,4}$, V A Oboznov$^{3,4}$, S~V~Egorov$^{3,4}$, V O Shkolnikov$^{4,5}$, M J Wolf$^{\,6}$, D Beckmann$^{6}$, and A V Ustinov$^{1,2,3}$}

\address{$^1$ Physikalisches Institut, Karlsruhe Institute of Technology, Wolfgang-Gaede-Str. 1, D-76131, Karlsruhe, Germany}
\address{$^2$ National University of Science and Technology MISiS, Leninskiy prosp. 4, Moscow, 119049, Russia}
\address{$^{3}$ Russian Quantum Center, Novaya str. 100, BC "Ural", Skolkovo, Moscow region, 143025, Russia}
\address{$^{4}$ Institute of Solid State Physics, Russian Academy of Sciences, Chernogolovka, Moscow region, 142432, Russia}
\address{$^5$ Moscow Institute of Physics and Technology (State University), Institutskiy per. 9, Dolgoprudny, Moscow region, 141700, Russia}
\address{$^6$ Institut f{\"u}r Nanotechnologie, Karlsruhe Institute of Technology, 76021 Karlsruhe, Germany}

\ead{ustinov@kit.edu}
\begin{abstract}
We describe fabrication and testing of composite flux qubits
combining Nb- and Al-based superconducting circuit technology. This
hybrid approach to making qubits allows for employing $\pi$-phase shifters fabricated using well-established Nb-based technology
of superconductor-ferromagnet-superconductor Josephson junctions.
The important feature here is to obtain high interface
transparency between Nb and Al layers without degrading  sub-micron shadow mask.
We achieve this by in-situ Ar etching using
e-beam gun. Shadow-evaporated Al/AlO$_x$/Al Josephson junctions with
Nb bias pads show the expected current-voltage characteristics with
reproducible critical currents. Using this technique, we fabricated
composite Nb/Al flux qubits with Nb/CuNi/Nb $\pi$-shifters and
measured their magnetic field response. The observed offset between
the field responses of the qubits with and without $\pi$-junction is
attributed to the $\pi$ phase shift. The reported approach can be
used for implementing a variety of hybrid Nb/Al superconducting
quantum circuits.
\end{abstract}

\pacs{03.67.Lx, 74.50.+r, 85.25.Cp} 

\vspace{1cm}
\noindent
{\it Keywords}: Superconducting qubits, Josephson junctions, flux qubits, pi-junctions, superconductor-ferromagnet junctions

%\submitto{\JPA}
% Comment out if separate title page not required
\maketitle

\section{Introduction}

One of several successfully implemented superconducting quantum circuits is a flux
qubit \cite{Mooij-1999,Chiorescu-2003}, which consists of a superconducting 
ring interrupted by three or four Josephson tunnel junctions. 
Though any kind of superconductor can be, in principle,
taken to make a qubit, the longest coherence times for flux and
other types of qubits are achieved by using shadow-evaporated
aluminum as the superconducting material and naturally grown 
aluminum oxide on top of it as the tunnel barrier. 
Shadow two-angle evaporation of aluminum using a suspended
electron-beam resist mask was established over thirty years ago
\cite{Niemeyer-1974,Dolan-1977} and is presently the most reliable
and widely used process for making sub-micron Josephson junctions. In
recent years, great progress has been achieved in applying this
junction manufacturing technique for superconducting quantum
circuits. Two-angle evaporation process has been successfully used
for a variety of superconducting qubit types
\cite{Clarke-Wilhelm-2008,You-Nori-2011} (charge, flux, transmon,
fluxonium, etc.) and appears to be most suitable for obtaining
well-defined sub-micron Al/AlO$_x$/Al Josephson junctions with
reliable characteristics and low density of microscopic two-level
defects in the oxide tunnel barrier
\cite{Lupascu-2009,Martinis-QIP-2009}. While niobium serves as the
base material for most of conventional superconducting circuits
employing Nb/AlO$_x$/Nb Josephson junctions, quantum coherence times
of Nb-based qubits
\cite{Oliver-2007,Lisenfeld-2007,Buisson-Zorin-2009} are
significantly shorter than those of their Al-based counterparts.
Aluminum superconducting flux qubits can be made very compact, while
well-controlled sizes of Josephson junctions defined by two-angle
evaporation make it possible engineering qubit potential with
precisely defined parameters \cite{Jerger-EPL-2011}.

The magnetic bias needed to drive the flux qubit to its working point
is a source of significant noise, leading to dephasing. The flux
qubit has the most favorable operation point with minimal dephasing
at the value of magnetic flux threading its loop of about
$\Phi_0/2$, where $\Phi_0$ is the magnetic flux quantum. In order to
reduce the effects of external magnetic noise, it was proposed to
avoid magnetic biasing by using the so-called $\pi$-junction in the
qubit loop \cite{Ioffe-1999,Blatter-2001}. The most reliable and
well-established process of implementing $\pi$-junctions relies on
Nb-based technology of superconductor-ferromagnet-superconductor
(SFS) Josephson junctions \cite{Oboznov-PRL-2006}. The conventional
fabrication process of $\pi$-junctions is based on the depositing of
an SFS Nb/CuNi/Nb trilayer, forming the junction, followed by
depositing the upper Nb wiring \cite{Oboznov-PRL-2006}. A
$\pi$-junction in the superconducting loop having large enough
critical current acts as a phase battery which biases the loop in the way
that the phase shift on the junction is $\pi$ \cite{Frolov-2008}.
The effect of such a phase shifter is equivalent to applying flux of
$\Phi{_0}/2$ through the loop \cite{Frolov-2008,Ust-Kap-2003}. SFS
phase shifters have already been successfully implemented in
Nb-based phase qubit circuits \cite{Feofanov-NatPhys-2010}, but they 
haven't yet been used in flux qubits. A complication arising along this
development is to combine two completely separated and not easily compatible
technological steps, first one for making relatively large SFS
junctions based on Nb, followed by another process of manufacturing
more fragile sub-micron Al junctions needed for highly-coherent flux
qubits. Aluminum two-angle evaporation is performed in a separate setup,
and pre-fabricated Nb structure is exposed to the air under which
the natural oxide NbO$_x$ is formed on Nb surface. This complication
makes the implementation of the SFS $\pi$-junction in the Al flux
qubit loop challenging and requires removing the NbO$_x$ completely
before the deposition of Al part of the flux qubit.

This paper is organized as follows. Section 2 describes the
combined Nb/Al technology for preparation of $\pi$-qubits. 
In Section 3, we present current-voltage measurements of Al/AlO$_x$/Al 
Josephson junctions with electrodes deposited on Nb pads. The developed
technology allows for obtaining high quality Al/AlO$_x$/Al Josephson
junctions  without residual resistance
at Nb/Al interface. In Section 4, we present measurements of flux qubits 
with Nb/CuNi/Nb $\pi$-junctions.

\section{Fabrication}

\begin{figure}[h]
\centering
\includegraphics[width=0.7\linewidth]{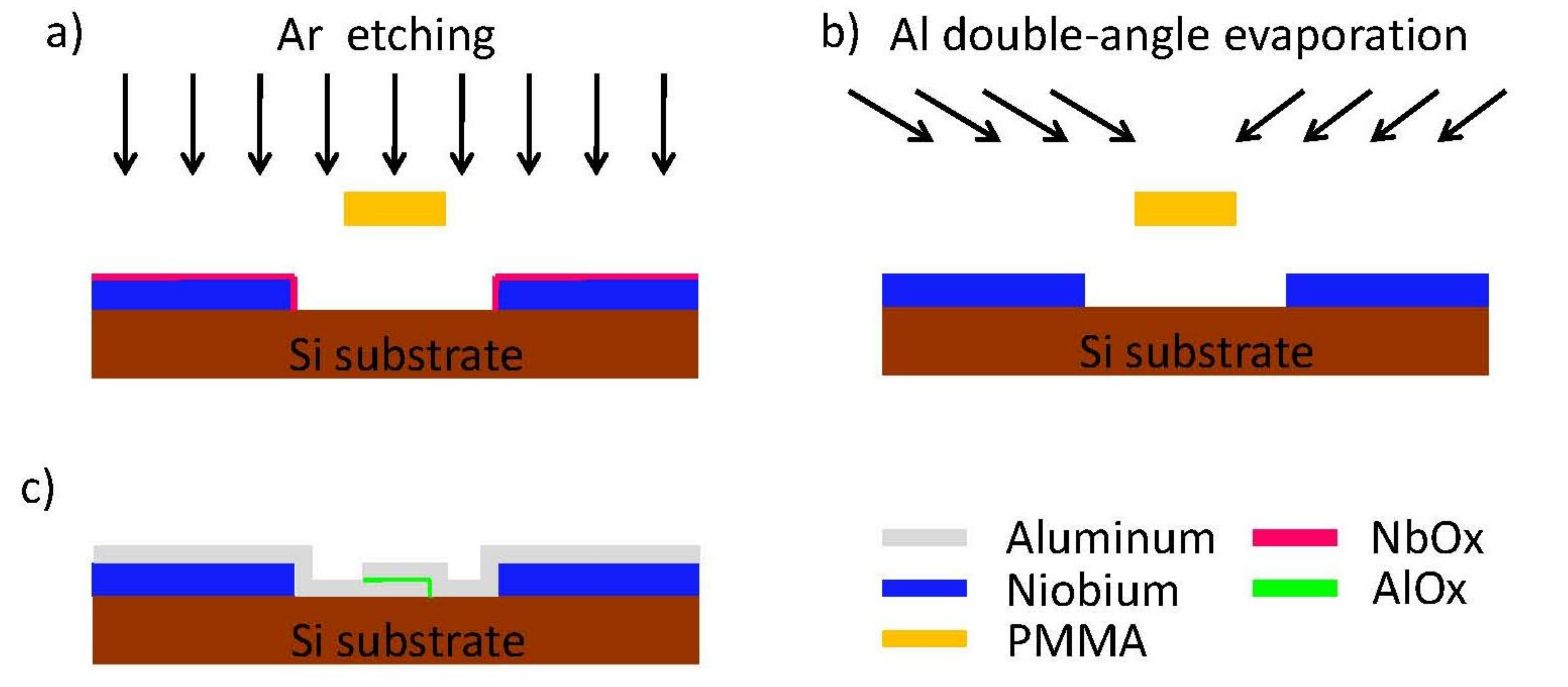}
\caption{Fabrication process of Al/AlO$_x$/Al Josephson junctions on Nb pads. (a) Etching of NbO$_x$ with the directed Ar beam in-situ right before the deposition of Al. (b) Double-angle Al shadow evaporation. (c) Resulting structure of Al Josephson junction on Nb pads free from NbO$_x$ after the lift-off procedure.}
\label{fig:TechP6}
\end{figure}

As the starting point, we describe our fabrication process of Al/AlO$_x$/Al
Josephson junctions with the Nb contact pads. Schematically, the
process is shown on Fig. \ref{fig:TechP6}. Prior applying this process, Nb
pads are fabricated in a separate vacuum chamber. The main
difficulty for achieving a good superconducting contact between Nb
and Al here is caused by a layer of non-superconducting NbO$_x$,
formed on the surface of Nb due to exposure to the air between the two processes.

The Nb contact pad layer was deposited by DC magnetron sputtering
and its patterning was done with the help of conventional optical
lithography. After that the sample was covered by a double-layer
resist and exposed in an electron-beam lithography machine to define
the desired structure for the following double-angle evaporation of
Al. Upon transferring the sample to Al deposition chamber, the
surface layer of NbO$_x$ was etched away in-situ using the directed
Ar beam in order to create a clean Nb surface before deposition of
Al. The specific  data for interface transparency after Ar-cleaning of
Nb-surface could be found in Ref.\cite{Oboznov-PRL-2006}, for
example.

Several measures for the etching procedure were taken aiming at
preventing the resist pattern from melting. We pre-cooled the
sample for 1 hour in a main chamber of evaporation machine at a high
vacuum of 10$^{-9}$ mbar at the temperature of about
T$\,\approx\,$-120 $^\circ$C. The layer of NbO$_x$ was etched away
in the load-lock by the directed Ar beam in 4 periods of 30 seconds
each, interrupted by 1 minute pauses. After that, the aluminum
Josephson junctions were deposited using the standard double-angle
shadow evaporation and oxidation \cite{Niemeyer-1974,Dolan-1977}. We
found out that the Ar beam current density of 20 $\mu$A/cm$^2$ was
sufficient to etch down the NbO$_x$ layer and establish the
superconducting contact between Al and Nb. The whole fabrication
process is shown in Fig. \ref{fig:TechP6}.

We fabricated a series of Al/AlO$_x$/Al Josephson junctions having
the dimensions 0.2$\times$1.0 $\mu$m$^2$. Furthermore, the flux qubits
interrupted by three Josephson junctions of this kind were
fabricated. The dimensions of Josephson junctions in flux qubits were
0.2$\times$0.5 $\mu$m$^2$ for the two junctions and 0.2$\times$0.335 $\mu$m$^2$
for the smaller $\alpha$-junction aiming at $\alpha$=0.67 \cite{Mooij-1999,Chiorescu-2003}.

\section{Characterization of Al/AlO$_x$/Al Josephson junctions}

\begin{figure}[h]
\centering
\includegraphics[width=0.7\linewidth]{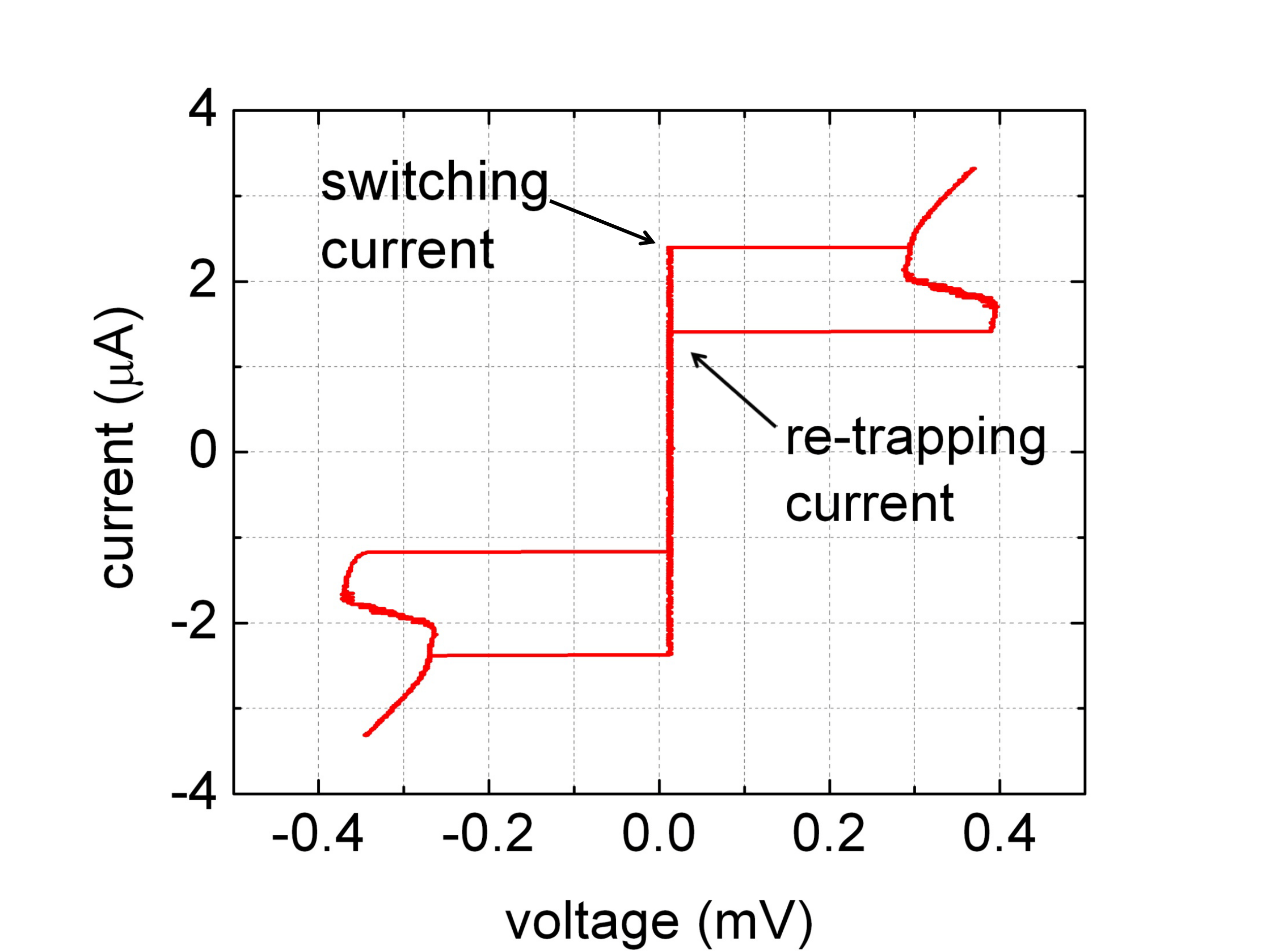}
\caption{Typical current-voltage characteristics of a hybrid Nb/Al/AlO$_x$/Al/Nb Josephson junctions having the dimensions 0.2$\times$1.0 $\mu$m$^2$. The re-trapping current "back-bending" is explained by the self-heating. }
\label{fig:IVmultiple3}
\end{figure}

Fabricated samples were measured in vacuum attached to a sample
holder of a He-3 cryostat and cooled down to a temperature of 300
mK. The current-voltage characteristics were measured using a
four-point configuration in the current-bias mode. Figure
\ref{fig:IVmultiple3} shows typical IV-curve for one of the test Al/AlO$_x$/Al
junctions having the dimensions 0.2$\times$1.0 $\mu$m$^2$. One can see a clear supercurrent branch as high as 2.5~$\mu$A. Switching
current values of 2 $\pm$ 0.54 $\mu$A and re-trapping currents of
around 1 $\pm$ 0.25 $\mu$A were measured for junctions made on
several chips. According to the process described above, all the contact pads
of our junctions are made of niobium so that each lead contained
Nb/Al interface. The current-voltage characteristics of Al/AlO$_x$/Al
Josephson junctions showed no evidence
of any residual resistance due to Nb/Al interfaces. 
So thus the quality of the fabricated Nb/Al/AlO$_x$/Al/Nb hybrid structures is high enough for implementation in quantum circuits.

From the topology of the junctions we estimated the junction capacitance
C$\,\approx\,$4.43$\pm$ 0.92 fF. The pronounced "back-bending" of
the re-trapping current branch visible in Fig. \ref{fig:IVmultiple3}
is rather typical for vacuum-based transport measurements of small
aluminum junctions in this range of the critical current density and
can be explained by non-equilibrium effects due to the
junction self-heating. Same phenomenon is also responsible for about 20$\%$
reduction in the measured value of the gap voltage V$_g$
\cite{Lotkov-2006}.

 \section{Flux qubit measurements}

\begin{figure}[h]
\centering
\includegraphics[width=0.7\linewidth]{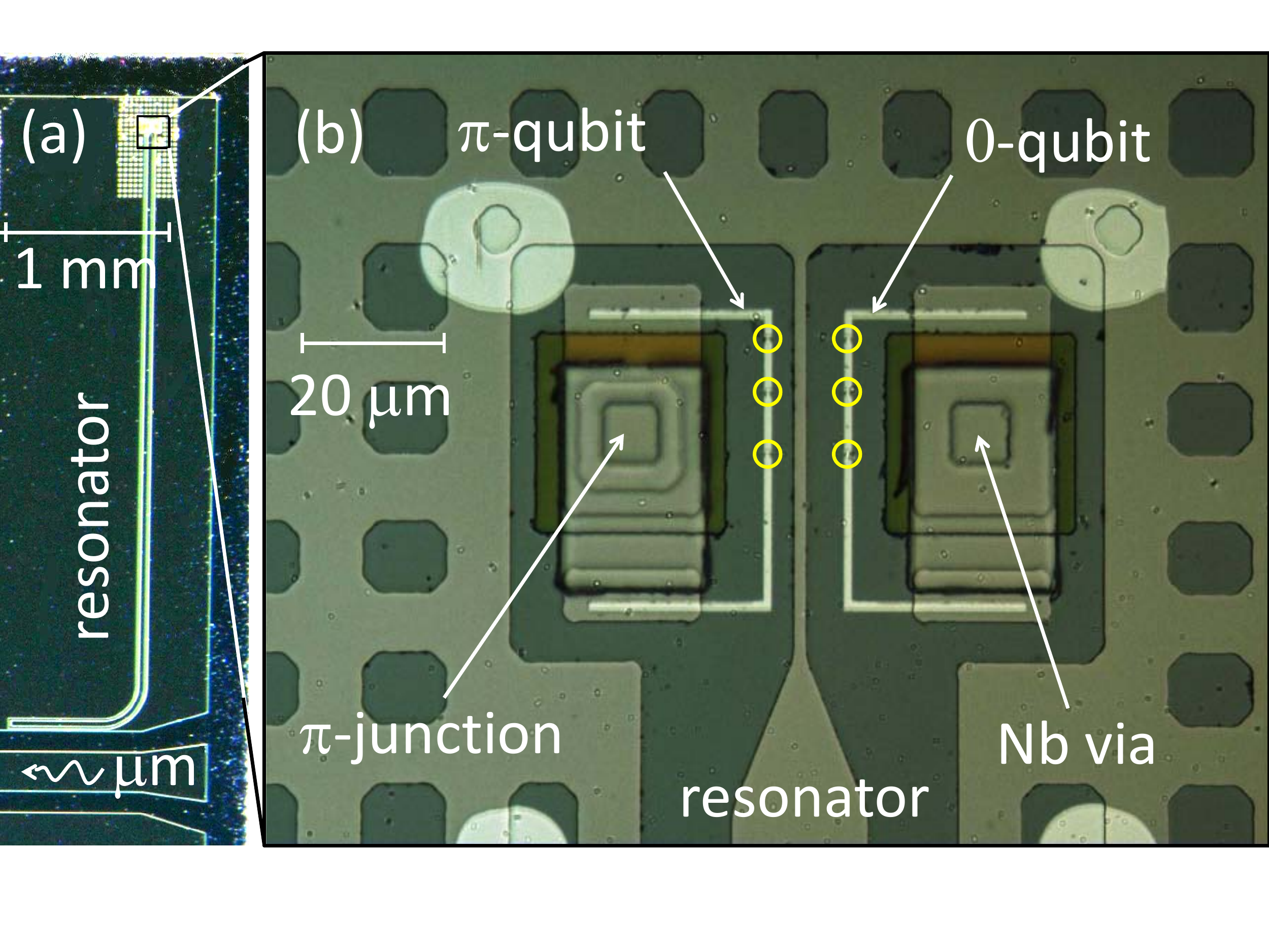}
\caption{ (a) $\lambda$/4 resonator capacitively coupled to the transmission line. (b) Optical picture of the two composite Nb/Al flux qubits placed near the shorted end of the $\lambda$/4 resonator. Nb part of the left qubit contains the $\pi$-junction. Right qubit has Nb "via" structure forming a superconducting short. The circles mark the positions of aluminum Josephson junctions. }
\label{fig:Design}
\end{figure}

For measurements of the flux qubits we used the conventional
dispersive readout setup discussed in detail elsewhere
\cite{Jerger-EPL-2011}. Two flux qubits were placed near the shorted
end of the $\lambda$/4 resonator, one with the SFS $\pi$-junction
with 12 nm layer of Cu$_{0.47}$Ni$_{0.53}$ \cite{Oboznov-PRL-2006},
and another one without it. SFS $\pi$-junction fabrication
technology is described in detail in \cite{Bol'ginov-2012}. The
opposite open end of the resonator was capacitively coupled to an
on-chip coplanar waveguide.  The micrograph, shown in
Fig.~\ref{fig:Design}, illustrates the sample.

The magnetic flux through the qubit loops was applied by using an
external magnetic bias coil. Measurements were performed in a
dilution cryostat at the base temperature of 25 mK.  We swept electrical
current through the bias coil, thus changing the flux through the
qubit loops. Simultaneously, the resonator was probed at its
fundamental $\lambda$/4 mode frequency $\omega_{r}$ with a microwave
signal transmitted and detected via a vector network analyzer (VNA).
We measured the amplitude and phase responses of the probe signal
amplified with a low-noise cryogenic amplifier at a fixed frequency
$\omega_{r}/(2\pi)=10.218$ GHz.

In the employed dispersive readout scheme, the resonator acquires a
dispersive shift due to the coupling to the qubit
\cite{Blais-2004,Wallraff-2004}
\begin{equation}
\Delta\omega_{r}$=$\pm$ $\frac{\tilde{g}}{\omega_{q}-\omega_{r}}\;,
\label{eq:DispersiveShift}
\end{equation}
where $\tilde{g}$ is an effective coupling of the resonator to the
qubit, $\omega_{q}$ is the transition frequency between the
$\mid$0$>$ and $\mid$1$>$ qubit states, $\omega_{r}$ is the resonant
frequency of the unperturbed resonator. From
Eq.~(\ref{eq:DispersiveShift}) one can see that, for the qubit
far-detuned from the resonator frequency, the dispersive shift is
small. When the qubit frequency $\omega_{q}$ approaches the
resonator frequency $\omega_{r}$, a relatively large dispersive
shift $\Delta\omega_{r}$ occurs.

For the flux qubit with the $\pi$-junction, we made the
critical current of the $\pi$-junction much larger than the
critical current that of any Al/AlO$_x$/Al Josephson junctions forming the
qubit. The following parameters have been used for
Nb/Cu$_{0.47}$Ni$_{0.53}$/Nb $\pi$-junction: CuNi-layer thickness of
12~nm, critical current density of 3.7 kA/cm$^2$ \cite{CNT2014}, mesa
size 10$\times$10 um$^2$, and the estimated critical current of about 3.7 mA. The value of
SFS-junctions critical current was about three orders
in magnitude larger than for tunnel junctions. 
In this case, due to relatively small persistent current flowing in the qubit loop, the
phase difference across the $\pi$-junction remains always close to
$\pi$, even at zero magnetic field. 
That causes the phase drop across aluminum tunnel junctions of the qubit to be shifted by the value of $\pi$, that is
equivalent to applying flux $\Phi_0$/2 in the qubit loop. 
Two qubits in our
experiment can be distinguished and measured
simultaneously with a single resonator because of their slightly
different loop areas and coupling strengths ($\tilde{g_1}$ and
$\tilde{g_2}$) to the resonator, provided by different
distances between the qubit loops and the resonator wire, see
Fig.~\ref{fig:Design}.

\begin{figure}[h]
 \centering
 \includegraphics[width=0.8\linewidth]{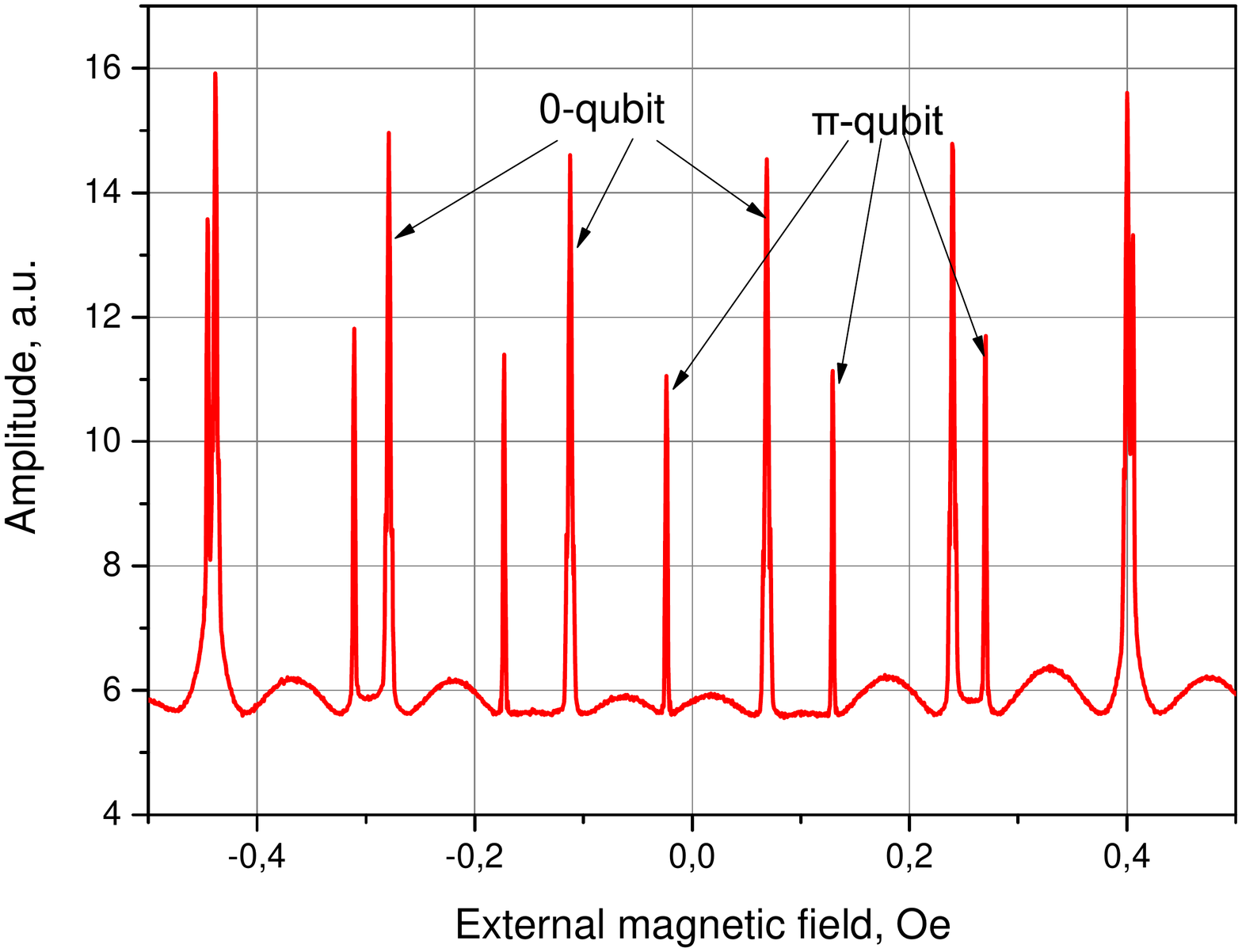}
 \caption{Amplitude of the dispersive response measured with a probe signal through a transmission line at a fixed frequency $\omega=\omega_{r}=10.218$ GHz. A periodic pattern with peaks of larger amplitude  corresponds to the flux qubit without $\pi$-junction, while smaller peaks are referred to the flux qubit with $\pi$-junction. The difference in amplitudes of the signals is attributed to the different coupling and detuning of the qubits from the resonator.}
 \label{fig:FieldResponse4}
 \end{figure}

The field response of the resonator coupled to two flux qubits, one with
and another without $\pi$-shifter is shown in Fig. \ref{fig:FieldResponse4}.
A peak in the transmitted microwave amplitude occurs when the
frequency of the transition between the ground and excited state of one of either qubit $\omega_{q1}$ or $\omega_{q2}$
approaches the resonator frequency $\omega_{r}$, which occurs at the
magnetic flux values close to $\Phi_0/2\pm n\Phi_0$ for the qubit without
$\pi$-junction and at $\pm n\Phi_0$ for the qubit with
$\pi$-junction, where $n$ is an integer.

\begin{figure}[h]
\centering
\includegraphics[width=0.8\linewidth]{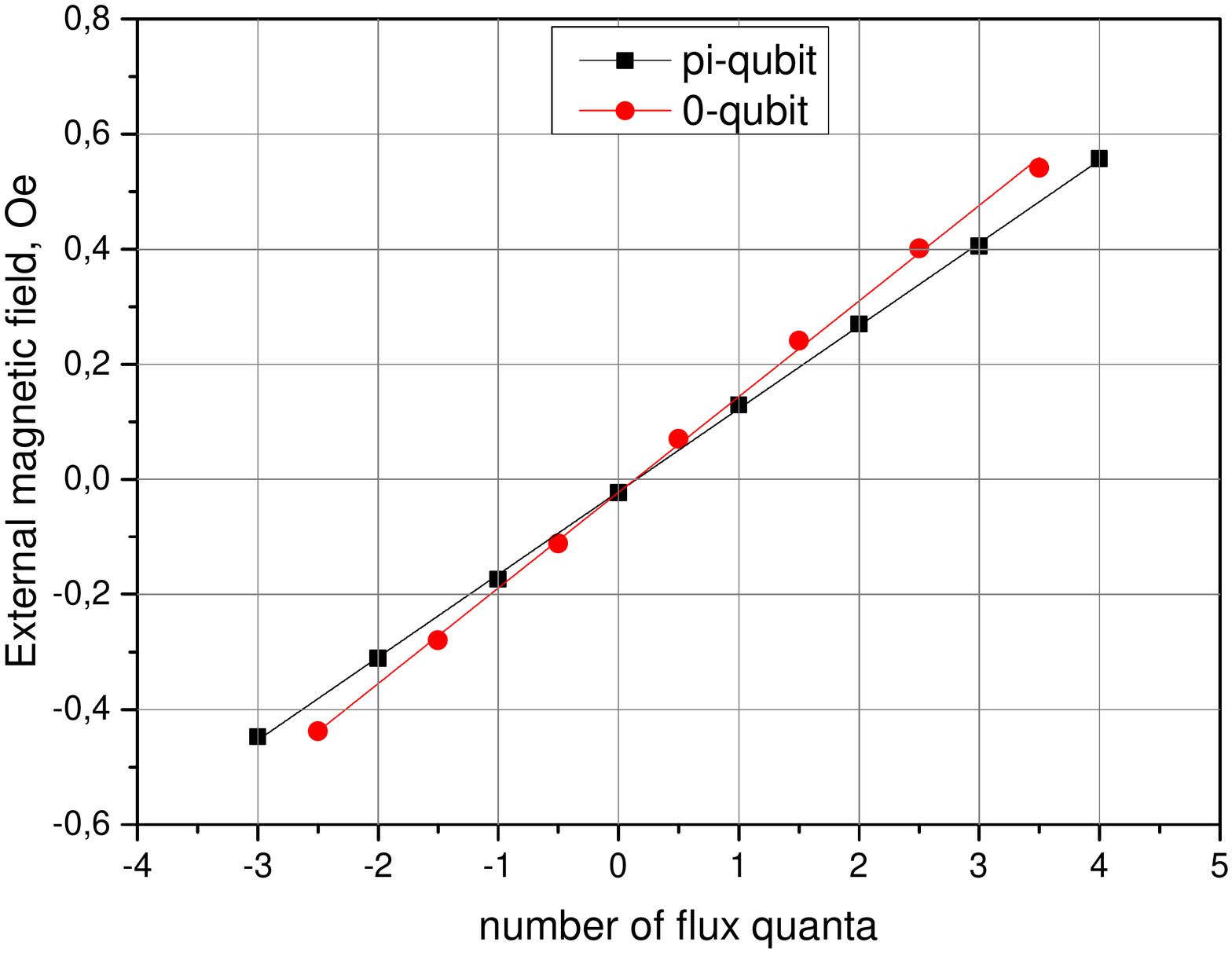}
\caption{The magnetic field bias vs flux quanta per qubit loop, extracted from positions of peaks in Fig. 4. The horizontal axis offset is chosen to have peaks of $\pi$-qubit at integer values of $\Phi/\Phi_0$.}
\label{fig:DistinguishPiNonPi}
\end{figure}

Next, we need to sort out two  families of periodic peaks in Fig. \ref{fig:FieldResponse4}, one of them
corresponding to 0-qubit and another to $\pi$-qubit. This
procedure is not straightforward because of non-ideal magnetic shielding.
Indeed, one can see from Fig. \ref{fig:FieldResponse4} that there is
no peak exactly at zero magnetic field, indicating the presence
of residual magnetic field in the setup. In Figure \ref{fig:DistinguishPiNonPi}, we have
plotted the positions of peaks as a function of magnetic flux. We
assumed that a period for each peak family is one flux quantum,
the residual flux of less than one flux quantum and the nearest-to-zero peak
as corresponding to $\pi$-qubit. One can see that both peak families could be
approximated by linear dependencies, each having its own slope, and
the intersection between two families takes place at zero net magnetic field.
An intersection point has to correspond to zero magnetic
flux, at which the $\pi$-qubit should display here a peak in the dispersive signal. 
Under this assumption, the residual magnetic field is equal to approximately 0.02
Oe and the residual magnetic flux is less than one flux quantum. 
One can easily identify two periods of oscillations in Fig. \ref{fig:FieldResponse4}.  
We suppose that the smaller period of the $\pi$-junction qubit oscillations in the applied 
magnetic field can be associated with an additional Josephson inductance of its loop 
induced by the $\pi$-junction. The smaller peak amplitude of the $\pi$-junction qubit 
response can be related to a larger detuning of its gap frequency from the resonator frequency, 
as well as to the additional inductance mentioned above. 
The energy gap of a flux qubit is extremely hard to control due to its very sensitive dependence on the relation between critical currents for three aluminum junctions. 

It is important to discuss the uniqueness of $\pi$-qubit response identification.
Actual values of magnetic flux in Fig.
\ref{fig:DistinguishPiNonPi} are assigned with possible offset by an integer
number of flux quanta. However, this circumstance doesn't alter our
definition of 0- and $\pi$-qubit responses. Indeed, an offset in definition of
zero-flux peak shifts \textbf{all} points of \textbf{both} peak families
in Fig. \ref{fig:DistinguishPiNonPi} along the horizontal axis by
an integer value. In this case, the crosspoint flux value will
change but it will remain to be integer-valued and the same peak
will correspond to the crosspoint. While our definition of zero magnetic
flux is just an assumption, the identification of $\pi$-qubit peaks base 
on the above described arguments seems unambiguous.

 \section{Conclusion}

In conclusion, we developed a fabrication process which allows to
create a superconducting interface between Nb and Al thin films produced 
in different technological processes. The developed procedure features pre-cooling
and Ar etching procedure of oxidized Nb surface in-situ, 
before deposition of Al/AlO$_x$/Al  Josephson
junctions using a standard double-angle shadow evaporation. This
process can also be implemented for more complex Nb/Al qubit
circuits. Hybrid Al/AlO$_x$/Al  flux qubits containing Nb/Cu$_{0.47}$Ni$_{0.53}$/Nb 
$\pi$-shifters were fabricated using the developed approach. We observed the field response 
of two flux qubits (one with and another without
$\pi$-shifter) coupled to the same $\lambda$/4 resonator. The
magnetic field shift between two periodic qubit oscillation patterns
measured at mK temperatures indicates the expected $\pi$-junction
phase bias in one of the flux qubit loops. The use of
the $\pi$-shifter makes it possible to avoid magnetic biasing, normally needed 
for reaching the most favorable flux qubit operating point, and thus to reduce
unavoidable variations of magnetic bias between different qubits on chip.

 \section*{Acknowledgments}

This work was supported in part by the  Russian Quantum Center,  the Ministry of Education and Science of the Russian Federation under contract no. 11.G34.31.0062 and 
no. К2-2014-025 (in the framework of Increase Competitiveness Program of  NUST MISiS), the Programs of the Russian Academy of Sciences, the Deutsche Forschungsgemeinschaft (DFG) and the State of Baden-W{\"u}rttemberg through the DFG Center for Functional Nanostructures (CFN).

\end{document}